\documentclass[letterpaper,11pt]{article}
\usepackage{jheppub}
\usepackage{physics}
\usepackage{relsize}
\usepackage{tikz}
\usepackage{pgfplots}
\usepackage{comment}
\usepackage{appendix}
\usepackage{esint}
\usepackage[utf8]{inputenc}
\title{$T\bar{T}+\Lambda_2$ Deformed CFT on the Stretched dS$_3$ Horizon}
\author{Vasudev Shyam}
\affiliation{Stanford Institute for Theoretical Physics and Department of Physics, Stanford University, Stanford, CA 94305, USA}
\date{\today}
\emailAdd{vshyam@stanford.edu}

\begin{document}
\abstract{
The $T\bar{T}+\Lambda_2$ deformation of a Holographic CFT on a cylinder (or torus in Euclidean signature) inhabits the stretched cosmological horizon in dS$_3$ when the deformation parameter is tuned to a particular value. I will describe how this insight allows us to compute the entropy of the stretched cosmic horizon including the logarithmic correction. 
}
\maketitle

\section{Introduction}
The holographic principle is the hypothesis that equates quantum gravitational physics in some spacetime volume to the physics of a quantum theory on the boundary of that region \cite{Susskind_1995, Stephens_1994}. This principle can be seen as an extrapolation of the observation that the entropy of black holes scale with the area of their event horizon, hinting that there is a localization of degrees of freedom relevant to the entropy to the horizon itself. This observation can then be promoted to bound the entropy of quantum fields in arbitrary regions of space \cite{Bek_bound, Bousso_1999}. Following this line of thought, a complete realisation of the holographic principle comes by the way of a duality that \emph{equates} quantum gravitational physics in some region to the physics of a quantum field theory inhabiting its boundary. The most well studied example of this duality is the AdS/CFT correspondence, which maps quantum gravity in spaces with AdS asymptotics to a conformal field theory inhabiting the asymptotic boundary \cite{Maldacena_1999}. The AdS/CFT correspondence forms a template for how the holographic principle ought to be realised in more general settings, i.e. in regions of finite spacetime volume like the interior of a black hole horizon or the cosmological horizon in a space of positive cosmological constant. This means that the quantum gravitational physics within finite regions is still expected to be dual to a quantum field theory living on the (appropriately defined) boundary of this region, but this quantum field theory is not a conformal field theory.

Recently, we have learned that solvable, irrelevant deformations of two dimensional conformal field theories provide a Holographically dual description of gravitational physics in regions of finite bulk volume. In particular these bulk regions are AdS$_3$ with the asymptotic boundary replaced with a finite cutoff surface \cite{MMV} and a region bounded by a timelike surface of constant radius in dS$_3$ \cite{dSdSTTbar}. In the former case, the dual theory inhabiting the cutoff surface is the $T\bar{T}$ \cite{Smirnov:2016lqw,Cavaglia:2016oda} deformation of the Holographic CFT that would otherwise inhabit the asymptotic boundary. In the latter case, the theory living on the constant radius surface in dS$_3$ is the combined $T\bar{T}+\Lambda_2$ deformation of a holographic CFT. Here $\Lambda_2$ denotes the cosmological constant of the boundary theory. This is the setting of interest for this article.

The flow triggered by these operators maps on to the radial development of the associated constant radius surfaces through the bulk, as dictated by Einstein's equations. In particular, when the bulk theory is classical (in the large $c$ limit of the theory on the boundary), the flow equations associated to these irrelevant operators map on to the radial-radial component of the bulk Einstein equations, also known as the radial Hamiltonian constraint. At finite $c$, we shall see that the associated flow equations can be mapped to the radial Wheeler de Witt equation in the bulk, which is the quantization of the radial Hamiltonian constraint. 
\subsection*{Flow equation for the thermal partition function}
The $T\bar{T}$ deformation of a two dimensional conformal field theory gives rise to one parameter family of quantum field theories whose generating function  satisfies:
\begin{equation}
\partial_{\mu}\log Z = -\int d^2x \langle T\bar{T}\rangle, \label{eq:defeq}
\end{equation}
where irrelevant operator $T\bar{T}$ is given by 
\begin{equation}
T\bar{T}(x):= T^{ij}(x)T_{ij}(x) - (T^i_i(x))^2.
\end{equation}
This deformation was introduced in \cite{Smirnov:2016lqw,Cavaglia:2016oda} and the energy spectrum of the deformed theory was determined in terms of the spectrum of the seed CFT through \eqref{eq:defeq}. The initial condition for this flow is set by the seed CFT:
\begin{equation}
Z(\mu=0)= Z_{CFT}.
\end{equation}

In \cite{dSdSTTbar,Lewkowycz:2019xse}, this deformation was generalized to the so called $T\bar{T}+\Lambda_2$ deformation
\begin{equation}
\partial_{\mu}\log Z = -\int d^2x \langle T\bar{T}\rangle -\frac{1}{\mu}\int d^2
x.
\end{equation}
In other words, it is the $T\bar{T}$ deformation with the addition of the cosmological constant at every step, which we call $\Lambda_2$. 
In particular, I will be interested in thermal partition function of a CFT$_2$ that lives on $\mathbb{T}^2$ by either the $T\bar{T}$ deformation or by the combined $T\bar{T} +\Lambda_2$ deformation. The flow equation for the partition function is: 
\begin{equation}
 \left(\frac{\tau_2}{2}\left(\partial^2_{\tau_1}+\partial^2_{\tau_2}\right)+\left(\lambda \partial_{\tau_2}-\left(\frac{\lambda}{\tau_2}-1\right)\right)\partial_{\lambda}-\frac{\tau_2(1-\eta)}{2\lambda^2} \right)Z(\lambda,\tau_1,\tau_2)=0.\label{eq:fle}
\end{equation}
Here $\lambda$ is a $(-1,-1)$ modular form that is related to the deformation parameter through $\lambda = \mu/R^2$. The real and imaginary parts of the torus modulus are denoted by $\tau_1$ and $\tau_2$ respectively. 
If $\eta=1$ then this is the flow equation of a $T\bar{T}$ deformed CFT. This equation was derived in \cite{Datta:2018thy}. If $\eta =-1$ then this is the flow equation for the $T\bar{T}+\Lambda_2$ deformation of a CFT.

\subsection{$T\bar{T}$ holography in AdS$_3$ and the dS$_3$ generalization.}
The connection between holography in finite regions and $T\bar{T}$ deformed CFTs was first proposed in \cite{MMV}. It was shown that gravity in AdS$_3$ with a Dirichlet wall at a finite distance from any point in the interior is dual to a holographic CFT deformed by the $T\bar{T}$ operator living on the Dirichlet wall.

In particular, the authors of \cite{MMV} considered General Relativity in 2+1 dimensions with a negative cosmological constant:
\begin{equation}
S=\frac{1}{16 \pi G} \int d^{3} x \sqrt{-g_{3}}\left(R+\frac{2}{\ell^2}\right)-\frac{1}{8 \pi G} \int_{B} d^{2} x \sqrt{-g}(K+1)
\end{equation}
where the boundary $B$ is a finite, radial cutoff surface. In particular, the datum that is specified is the metric on $B$:
\begin{equation}
ds^2\vert_{B} = g_{ij} d x^{i} d x^{j}=-N^{2} d t^{2}+e^{2 \varphi}(d \theta-\omega d t)^{2}.
\end{equation}
This parameterization allows us to simply characterize the quasi local energy and momentum of the truncated spacetime. On shell, the variation of the action is given by:
\begin{equation}
\delta S=\int_{B} d^{2} x \pi^{ij} \delta g_{ij}=\int_{B} d^{2} x \sqrt{-g}(-\epsilon \delta N-j \delta \omega+p \delta \varphi)
\end{equation}
where $\epsilon,j$ and $p$ are the energy, momentum and pressure densities of the truncated spacetime. In particular, the quasi local energy of the theory is given by 
\begin{equation}
E^{ql} = -\frac{R}{G\ell}\left(-1+\sqrt{1-\frac{8 G M\ell}{R^2}+\frac{16 G^2 J^2}{R^4}}\right).
\end{equation}
This spectrum matches the energy levels of the $T\bar{T}$ deformed holographic CFT when we identify $c=3\ell/2G$ and $\mu = \frac{2G\ell}{\pi}.$ 

In \cite{dSdSTTbar}, it was shown that the gravitational physics within a finite patch of de Sitter space, particularly, the region $w<w_c$ of the geometry given by line element:
\begin{equation}
d s^{2}=d w^{2}+\sin^{2} \frac{w}{\ell} d s_{d S_{2}}^{2},
\end{equation}
is dual to the $T\bar{T}+\Lambda_2$ deformation of a holographic CFT living on the lower dimensional dS$_2$ surface of constant $w_c$. It was shown therefore that the $dS/dS$ correspondence could be realised through the $T\bar{T}+\Lambda_2$ deformation. Subsequently, in \cite{Lewkowycz:2019xse}, this correspondence was extended to other finite regions of dS$_3$, in particular to world tubes surrounding a static observer within their static patch. These world tubes are surfaces of constant $R$ in the geometry given by the following line element:
\begin{equation}
d s^{2}=\frac{d R^{2}}{\left(1-\frac{R^{2}}{\ell^{2}}\right)}+\left(1-\frac{R^{2}}{\ell^{2}}\right) d t_{E}^{2}+R^{2} d \phi^{2}.
\end{equation}
In the extreme limit $R\rightarrow \ell$ these world tubes approach the cosmic horizon. Strictly speaking, the world tubes approach the stretched cosmic horizon as $R$ approaches $\ell.$


\subsection*{Summary of the main result}
The main result of this article is the calculation of the log corrected de Sitter entropy formula in the bulk: 
\begin{equation}
S_{hor} = \frac{\pi\ell}{2G}-3\log \left(\frac{\pi\ell}{2G}\right)+\cdots
\end{equation}
as the logarithm of the density of states of a $T\bar{T}+\Lambda_2$ deformed holographic CFT, in the high temperature limit: 
\begin{equation}
S(\mathfrak{E},\bar{\mathfrak{E}},\lambda) =\log \rho (\mathfrak{E},\bar{\mathfrak{E}},\lambda)= \min_{\tau,\bar{\tau}}\bigg(i\frac{\tau \mathfrak{E}}{2}-i\frac{\bar{\tau}\bar{\mathfrak{E}}}{2}-\log Z(\tau,\bar{\tau},\lambda)\bigg).
\end{equation}
The key instrument to do this calculation is the flow equation that the density of states satisfies, that can be inferred from the $T\bar{T}+\Lambda_2$ flow equation that is satisfied by the partition function: 
\begin{equation}
\frac{\eta-1}{\lambda^2}(\partial_{\mathfrak{E}}\log \rho+\partial_{\bar{\mathfrak{E}}}\log \rho)- \mathfrak{E}\bar{\mathfrak{E}}(\partial_{\mathfrak{E}}\log \rho+\partial_{\bar{\mathfrak{E}}}\log \rho)+2\left(-1+\frac{\lambda(\mathfrak{E}+\bar{\mathfrak{E}})}{2}\right)\partial_{\lambda}\log \rho=0.
\end{equation}

\subsection{Overview of the calculation}
In section 2, I will show how a partition function written as 
\begin{equation}
Z = \sum_{n} e^{-2\pi \tau_2 \mathcal{E}_n(\lambda)+2\pi i \tau_i J_n },
\end{equation}
where the exponents contain the deformed energy levels $RE_{n}(\mu,R)=\mathcal{E}_n(\lambda = \mu/R^2)$ solve the flow equation, and I also show the explicit form of $\mathcal{E}_n(\lambda)$ in terms of the seed CFT data. Then I will show how to change variables to re-interpret these quantities in the bulk. 

In section 3, I will show what tuning of parameters in the deformed theory correspond to approaching the near horizon region of the bulk. 

Finally, in section 4, I will show how the density of states in the near horizon regime can be extracted from an inverse Laplace transformation of the generating function, and further what flow equation that quntity has to satisfy. Solving the flow equation will yield the Gibbons Hawking entropy of the cosmic horizon, along with the leading logarithmic correction. 
\section{Solving the flow equation}
In this section, the solution of the flow equation \eqref{eq:fle} in terms of the deformed energy levels will be discussed. Also, \eqref{eq:fle} is rewritten in terms of the radial Wheeler--de Witt equation in 2+1 dimensions with negative cosmological constant. 
\subsection{The energy spectrum}
I will be interested in the case where the CFT being deformed admits a large $c$ limit. However, the flow equation \eqref{eq:fle} can be solved at any value of $c$ (and $\eta$) via the ansatz: 
\begin{equation}
Z(\lambda, \tau_1,\tau_2) = \sum_n e^{-2\pi \tau_2 \mathcal{E}_n(\lambda)+2\pi i \tau_1 J_n}, \label{eq:psum}
\end{equation}
which is the expression of the partition sum, written in terms of dimensionless quantities. To recover a more familiar expression, we need to define:
\begin{equation}
\mathcal{E}_n(\lambda) = RE_n(R,\lambda),\,\,J_n = Rj_{n}(R).
\end{equation}
Here $E_n$ and $j_n$ are the dimensionful energy and angular momentum. We also notice that the inverse temperature $\beta$ is related to the imaginary part of the torus modulus:
\begin{equation}
\beta = R\tau_2. 
\end{equation} 
Since I am considering the deformation of a CFT, the partition function at $\lambda=0$ is given by: \begin{equation}
Z(\lambda=0,\tau_1,\tau_2) = Z_{CFT}(\tau_1,\tau_2) = \sum_n e^{-2\pi R E^o_n(R,\lambda=0)+2\pi i \tau_1 R j_n(R) }.
\end{equation}
Note that $j_n$ gains no $\lambda$ dependence due to the deformation, which is a reflects the fact that the $T\bar{T}+\Lambda_2$ deformation preserves rotation invariance along the spatial circle, i.e. the contractible cycle of the torus. The CFT quantities $E_n,j_n$ are given by: 
\begin{equation}
RE_n=\mathcal{E}^o_n=\Delta_n+\bar{\Delta}_n-\frac{c}{12},\,\, Rj_n = J_n = \Delta_n-\bar{\Delta}_n.
\end{equation}
Here $\Delta_n, \bar{\Delta}_n$ denote the left and right conformal dimensions of the conformal field theory, and $c$ denotes the central charge.

The equation \eqref{eq:fle} for the partition sum \eqref{eq:psum} implies the following equation for the deformed energy levels $\mathcal{E}_n(\lambda)$:
\begin{equation}
2\pi\lambda\mathcal{E}_{n} \partial_{\lambda}\mathcal{E}_{n}- \partial_{\lambda}\mathcal{E}_{n}+\pi\mathcal{E}^{2}_{n}+\frac{(1-\eta)}{4\pi\lambda^2} = \pi^2 J_n^2.
\end{equation}
This is the generalization of the Burgers' equation \cite{Smirnov:2016lqw,Cavaglia:2016oda} for the deformed energy levels first obtained in \cite{dSdSTTbar}. The solution for the deformed energy levels is given by: 
\begin{equation}
\mathcal{E}_n(\lambda) = -\frac{-1+\sqrt{\eta + 4\pi\left(-\lambda \mathcal{E}^o_n+2\pi\lambda^2 J^2_n\right)}}{\lambda}. 
\end{equation}

Note that this expression first appeared in \cite{Lewkowycz:2019xse}.

\subsection{Radial Wheeler--de Witt equation}
The flow equation \eqref{eq:fle} can be recast as the radial Wheeler de Witt equation (in a fixed gauge) in three dimensional quantum gravity. Following \cite{coleman2020conformal}, I first introduce the variable 
\begin{equation}
V = 4\pi^2 R^2 \tau_2 = 4\pi^2\mu\frac{\tau_2}{\lambda}.
\end{equation}
Here I have defined $\mu = R^2 \lambda$. Then, I define the radial wavefunction:
\begin{equation}
\psi(V,\tau_1,\tau_2)= e^{-\frac{V}{2\pi^2 \mu}}Z(V,\tau_1,\tau_2),
\end{equation}
that satisfies the radial WdW equation: 
\begin{equation}
16\pi G \ell \left(\frac{\tau^2_2}{2V}\left(\partial^2_{\tau_1}+\partial^2_{\tau_2}\right)-\frac{V}{2}\partial^2_{V}\right)\psi(V,\tau_1,\tau_2)+\frac{V\eta}{8\pi G\ell}\psi(V,\tau_1,\tau_2)=0.\label{eq:fincfle}
\end{equation} 
Note that $\eta$ indeed sets the sign of the cosmological constant. In the above, I have assumed that the Brown--Hennaux relation $c = \frac{3\ell}{2G}$ remains unchanged. We also need to identify 
\begin{equation}\mu = \frac{2 G \ell}{\pi},\end{equation}
in order to get the right constants in the WdW equation. 

We can check that this equation, (and therefore, its solutions) are modular invariant. In particular, the variable $V = 4\pi^2 \mu \frac{\tau_2}{\lambda}$ is modular invariant, and the eigen-functions of the Maass Laplacian: 
\begin{equation}
\Delta^{M}_o=\tau^2_2\left(\partial^2_{\tau_1}+\partial^2_{\tau_2}\right)
\end{equation}
are weight 0 modular forms, in other words, functions $f(\tau = \tau_1+i \tau_2,\bar{\tau}=\tau_1-i\tau_2)$ such that 
\begin{equation}
f\left(\frac{a \tau+b}{c \tau +d},\frac{a \bar{\tau}+b}{c \bar{\tau} +d}\right) = f(\tau,\bar{\tau}),
\end{equation}
where $a,b,c,d\in \mathbb{Z}$ and $ad-bc=1$. In the bulk gravity theory, modular transformations are the large diffeomorphism group acting on the constant radius surfaces. Modular invariance will be crucial to obtain the density of states and cosmological horizon entropy. 
\subsection{Semiclassical limit in the bulk}
The regime where we recover classical physics in the bulk is where we take $c\rightarrow \infty,\,\mu\rightarrow 0$ while holding $\mu c$ fixed. In terms of the bulk parameters, this is equivalent to taking $G\rightarrow 0$ while holding $\ell^2$ fixed. We can make the following ansatz for the solution:
\begin{equation}
\psi(V,\tau_1,\tau_2) = e^{-c\mathcal{W}(V,\tau_1,\tau_2)}
\end{equation}
and define the momentum operator as:
\begin{equation}
\hat{p}_{\tau_a}=-\frac{1}{c}\partial_{\tau_a},\,\,\hat{p}_{V}=-\frac{1}{c}\partial_{V}.
\end{equation}
In the large $c$ limit this gives us the following equation:
\begin{equation}
H=16\pi G\ell\left(-\frac{V}{2} p^2_{V}+\frac{\tau^2_2}{2V}\left(p^2_{\tau_1}+p^2_{\tau_2}\right)\right)+\frac{V\eta}{8\pi G\ell}=0, \label{eq:hc}
\end{equation}
where the classical momenta are defined as
\begin{equation}
p_{V}=\frac{1}{c}\frac{\partial_{V}\psi}{\psi}=\partial_{V}\mathcal{W}(V,\tau_1,\tau_2) \label{momv}
\end{equation}
\begin{equation}
p_{\tau_a}=\frac{1}{c}\frac{\partial_{\tau_a}\psi}{\psi}=\partial_{\tau_a}\mathcal{W}(V,\tau_1,\tau_2).\label{momp}
\end{equation}
The terms suppressed by $1/c$ are of the form $\partial^2_{V}\mathcal{W}$ and $\partial^2_{\tau_a}\mathcal{W}$. The equation \eqref{eq:hc} is the classical (radial) Hamiltonian constraint gauge fixed in Constant Mean Curvature gauge. This equation is quadratic in the first derivative with respect to Hamilton's principal function $\mathcal{W}$, so it is non linear - this in turn implies that the superposition that was the solution of the finite $c$ equation \eqref{eq:fincfle} will not solve this equation. However, one term in that sum does solve the equation. This corresponds to the classical geometry that solves the constraint equations in the bulk. We take: 
\begin{equation}
\mathcal{W}=\frac{1}{c}\left( 2\pi i J  \tau_1 + \frac{\tau_2}{\lambda}\left(\sqrt{\eta+4\pi\left(-\lambda \mathcal{E}+\pi J^2\lambda^2\right)}\right)\right).
\end{equation}
I have dropped the subscript $n$ because are are looking only at one state at a time. In order for this solution to contributes to the $c\rightarrow \infty$ limit, the weights have to scale with $c$. In other words
\begin{equation}
\mathcal{E} = c\left(\delta + \bar{\delta}-\frac{1}{12}\right)=c\bar{\mathcal{E}},
\end{equation}
\begin{equation}
J = c\left(\delta - \bar{\delta}\right)=c\bar{J},
\end{equation}
where $\delta, \bar{\delta}$ are order 1. Then, I will also define 
\begin{equation}
\lambda'= \frac{\pi\lambda c}{3}.
\end{equation}
In terms of these quantities, we have the following Hamilton's principal function: 
\begin{equation}
\mathcal{W}= 2\pi i \bar{J} + \frac{\pi\tau_2}{3\lambda'}\left(\sqrt{\eta +\left(-12\bar{\mathcal{E}}\lambda'+36\bar{J}^2\lambda'^{2}\right)}\right).
\end{equation}
To simplify the analysis and to make contact with previous work, I will consider the zero momentum sector:
\begin{equation}
\mathcal{W}= \frac{\pi\tau_2}{6\lambda'}\sqrt{\eta- 12\bar{\mathcal{E}}\lambda'}.
\end{equation}

Written in terms of the volume $V= 4\pi^2 \frac{\mu'}{\lambda'}$\footnote{$\mu' = \frac{\pi c \mu}{3}$}, we have:
\begin{equation}
\mathcal{W}= V\sqrt{\eta- 48\pi^2 \frac{\mu' \tau_2\bar{\mathcal{E}}}{V}}.
\end{equation}

\section{The de Sitter horizon}
The above equation leads to a formula for the quasi local energy in the bulk space-time, which is really a region of finite volume bound by a time-like surface whose location is set by $\mu'$. The relation between the quasi local energy $E^{ql}$ and $\mathcal{W}$ above is given by:
\begin{equation}
R E^{ql} = \frac{c \pi}{3\lambda'}+\frac{c}{\tau_2}\mathcal{W}.
\end{equation}
My interest will be in the square root singularity of the function $\mathcal{W}$ and how this corresponds to the location of the horizon in the bulk. 
\subsection*{$\mathbf{\eta=1}$}
In the case where $\eta =1$, the argument of the square root vanishes when 
\begin{equation}
\lambda' = \frac{1}{12 \bar{\mathcal{E}}} \label{btzhor}
\end{equation}
Note that for the ground state, $\bar{\mathcal{E}}=-1/12$ and this equation cannot be satisfied, because the $\lambda$ is always positive. Therefore, excited states with $\Delta \sim c $ see this singularity. These corresponds to BTZ black holes in AdS space. The condition above can be unfurled to show that it corresponds to putting the boundary on the horizon. In other words, it is the condition that the surface on which the theory lives hugs the horizon. In other words, it is the limit where the deformed theory inhabits the stretched horizon. The relation \eqref{btzhor} translates into $R = \sqrt{8GM\ell}$ which is a more recognisable relation for the location of the horizon\footnote{Here I use $RE = M$}.

\subsection*{$\mathbf{\eta=-1}$}
Here, the deformation of the ground state itself sees a square root singularity. The argument of the square root vanishes when
\begin{equation}
\lambda'=1\,,R^2 = \ell^2.
\end{equation}
This relation follows from taking $\bar{\mathcal{E}} = -\frac{1}{12}$. In Euclidean signature, from  looking at the $dS_3$ metric in the static patch 
\begin{equation}
ds^2 = \frac{dR^2}{\left(1-\frac{R^2}{\ell^2}\right)}+\left(1-\frac{R^2}{\ell^2}\right)dt_E^2+R^2 d\phi^2. 
\end{equation}
Note that the topology of the constant $r$ slices is that of $\mathbb{T}^2 = S^1\times S^1$. The two circles are parameterized by $\phi$ and the Euclidean time $t_E$. the latter is the non contractible cycle and the former is the contractible cycle. At the horizon, the non contractible cycle pinches, and the topology changes to that of a hemisphere\footnote{This observation was also made in \cite{Suneeta:2002nr}, and the quantum state of the stretched horizon was computed from Chern--Simons Theory in \cite{Govindarajan_2002}.}. This sets up the expectation that $\tau_2$ should go to zero as we approach the horizon as a function of $r$ (the analogue of "time" in the Euclidean setting), the radial position. This is important to bear in mind if we want to understand what happens to the partition function as we approach the horizon. Hamilton's equations relate $\tau_2$ to derivatives of the energy function in a non trivial way and this will be what the following section aims at clarifying. 
\subsection{Probing the near - horizon regime}
The classical equations of motion for $\tau_2$ follow from \eqref{eq:hc}. In particular
\begin{equation}
\partial_{r}\tau_2 = \partial_{\tau_2}H = \frac{\tau^2_2}{V}p_{\tau_2}.
\end{equation}
We can use the relation $p_{\tau_2}=\partial_{\tau_2}\mathcal{W}$ to rewrite the right hand side:  
\begin{equation}
\partial_{r}\tau_2= \frac{\tau^2_2}{V}\partial_{\tau_2}\mathcal{W}=-\frac{\tau^2_2}{V}\frac{ c \mathcal{E}\sqrt{V}}{\sqrt{12 \mu \tau_2\mathcal{E}+ \frac{V\eta}{4\pi^2}}}.
\end{equation}
We note that as a function of $V$, the near horizon regime is characterised by the behaviour of our quantities near the point $V= -\frac{48\pi^2\mathcal{E}\mu \tau_2}{\eta}=V_*$. The idea is to integrate the $\tau_2$ equation of motion near the point $V_*$. Doing this to leading order in the deviation $\epsilon$ away from $V_*$ we find:
\begin{equation}
\tau_2(r)= -\frac{3\epsilon}{4(c^2\pi^4 r^2\mathcal{E} \mu)}+O(\epsilon^{3/2}).
\end{equation}
This holds for either sign of $\eta$, but it relies on the fact that there is some singularity in the square root, which in turn occurs for different $\mathcal{E}$ values depending on the sign of $\eta$. I will focus in a later subsection on the $\eta = -1$ case where the ground state hits a square root singularity at some value of $V$ (or $\lambda$). The above result clarifies that in fact, as we approach the horizon, the non-contractible cycle of the torus shrinks to zero size. This implies: 
\begin{equation}
\psi(V,\tau_1,\tau_2)\vert_{V=V_*}\propto 1. 
\end{equation}

Furthermore, given that the translation between the wave-function $\psi$ and the partition function $Z$ is through multiplying by the phase $e^{-\frac{\tau_{2}}{\lambda}}$, we see that the same conclusion holds for $Z$:

\begin{equation}
Z(\lambda,\tau_1,\tau_2)\vert_{\lambda=\lambda_*}\propto 1. 
\end{equation}
The analysis of the equations of motion is inadequate to fix the proportionality factors, and that will be the aim of the subsection after the next. We do already see indications that the reduced density matrices for the system obtained from tracing out everything beyond the horizon in the bulk is that of a maximally mixed state. This result agrees with the expectation coming from the dS/dS correspondence \cite{Dong:2018cuv,Lewkowycz:2019xse}.  

Similarly, the partition function on the stretched horizon is given by: 
\begin{equation}
Z_{Stretched}=Z(\lambda = \lambda_*+\epsilon,\tau_1,\tau_2).
\end{equation}
This quantity is necessarily defined in the semi-classical regime, where it is meaningful to isolate an energy level and talk about the value of $\lambda$ at which the argument of the square root in the energy level vanishes. Then, by studying the $1/c$ expansion, we can understand some of the leading quantum effects affecting the horizon in the bulk. In particular, we can compute the energy and entropy associated to the horizon (either BTZ in the $\eta=1$ case, or the Cosmological horizon in the $\eta=-1$ case) from the partition function in the near horizon region, and extract the leading order in $c$ corrections to these results. 

\section{Density of states}
An interesting consequence of $\tau_2$ vanishing at the horizon is that the high temperature limit of the partition function is mapped to the near horizon limit in the bulk dual. The convenient thing about this coincidence is that the density of states can be extracted easily in the high temperature limit. Another property that I will exploit is the modular invariance of the deformed partition function. This will let us relate the high and low temperature expansions of the partition function.

Note that the exercise to follow is a de Sitter analog of the calculation of the horizon entropy of BTZ black holes through the Cardy formula of a dual CFT, as presented in \cite{Strominger_1998}, \cite{Carlip_2000}, and references therein. A similar analysis was performed to obtain the Kerr-- dS$_3$ horizon entropy from the CFT Cardy formula in \cite{Bousso_2002}. Their calculation assumes the dS/CFT correspondence, which is a different holographic dictionary than the one considered in this article.

The partition function can be written as: 
\begin{equation}
Z(\lambda, \tau, \bar{\tau}) = \int \int\, d \mathfrak{E}\,d \bar{\mathfrak{E}}\,\rho(\mathfrak{E},\bar{\mathfrak{E}}) e^{i\frac{\tau}{2} \mathfrak{E}}e^{-i\frac{\bar{\tau}}{2}\bar{\mathfrak{E}}}.\label{eq:Zerep}
\end{equation}
Here, $\rho(\mathfrak{E},\bar{\mathfrak{E}})$ is the density of states. It can be extracted from a double inverse Laplace transformations: 
\begin{equation}
\rho (\mathfrak{E},\bar{\mathfrak{E}})= \oiint\,d\tau \,d\bar{\tau} e^{-i\frac{\tau}{2} \mathfrak{E}}e^{i\frac{\bar{\tau}}{2}\bar{\mathfrak{E}}} Z(\lambda,\tau,\bar{\tau}).
\end{equation}
Note that $\mathfrak{E} = 2\pi \mathcal{E}$. 

I will be interested in studying the density of states in the high energy limit. This will be the method I will use to extract the density of states in the following subsection. 
\subsection{Density of states from the inverse Laplace transform}
\paragraph{The CFT case:}
For simplicity, I'll start with describing the calculation in the undeformed CFT. I will assume that this CFT is unitary, modular invariant and that it admits a large $c$ limit. The double inverse Laplace transform takes the form: 

\begin{equation}
\rho_{CFT}(\mathfrak{E},\bar{\mathfrak{E}})= \oiint\, d\tau d\bar{\tau} e^{-i\frac{\tau}{2} \mathfrak{E}}e^{-i\frac{\bar{\tau}}{2} \bar{\mathfrak{E}}}Z_{CFT}(\tau,\bar{\tau}).
\end{equation}
 
The high temperature limit of the partition function in a modular invariant CFT is given by the S transformation of the ground state:
\begin{equation}
Z_{CFT}\left(-\frac{1}{\tau},-\frac{1}{\bar{\tau}}\right)\vert _{\tau,\bar{\tau}\ll1} \simeq e^{i \frac{c\pi}{12 \tau}}e^{-i \frac{c\pi}{12 \bar{\tau}}}.
\end{equation}
In the case where the torus modulus is purely imaginary, 
\begin{equation}
Z_{CFT}\left(-\frac{1}{\tau_2}\right)\vert _{\tau_2\ll1,\tau_1=0}\simeq e^{-\frac{c\pi}{6\tau_2}}. 
\end{equation}
The high energy density of states therefore factorises: 
\begin{equation}
\rho_{CFT}(\mathfrak{E},\bar{\mathfrak{E}})= \rho_{CFT}(\mathfrak{E})\rho_{CFT}(\bar{\mathfrak{E}}),
\end{equation}
where:
\begin{equation*}
\rho_{CFT}(\mathfrak{E})= \oint\, d\tau e^{-i\frac{\tau}{2} \mathfrak{E}}Z_{CFT}\left(-\frac{1}{\tau}\right)\bigg\vert_{\tau\ll 1} = \oint d\tau e^{-i\left(\frac{\tau}{2} \mathfrak{E}- \frac{c \pi}{12 \tau}\right)},
\end{equation*}
\begin{equation}
\rho_{CFT}(\bar{\mathfrak{E}})= \oint\, d\bar{\tau} e^{i\frac{\bar{\tau}}{2}\bar{\mathfrak{E}}}Z_{CFT}\left(-\frac{1}{\bar{\tau}}\right)\bigg\vert_{\tau\ll 1}=\oint d\bar{\tau} e^{i\left(\frac{\bar{\tau}}{2} \bar{\mathfrak{E}}- \frac{c \pi}{12 \bar{\tau}}\right)}\label{eq:invlap}
\end{equation}
In the regime where $\tau,\bar{\tau}$ are small, these integrals can be evaluated in the saddle point approximation. Also, for the ground state, we have $\mathfrak{E}= \bar{\mathfrak{E}}$. Up to the first sub-leading order, \eqref{eq:invlap} evaluates to:  
\begin{equation}
\rho_{CFT}(\mathfrak{E}, \mathfrak{E})\simeq \frac{c^2\pi^3}{36 S^{3}_o(\mathfrak{E})}e^{S_o(\mathfrak{E})+O\left(\frac{1}{\mathfrak{E}}\right)}
\end{equation}
where 
\begin{equation}S_o(\mathfrak{E})= \sqrt{\frac{2c\pi\mathfrak{E}}{3}}.\end{equation} 

This is the famous Cardy formula and its leading logarithmic correction. The takeaway from this section is that the sub-leading corrections to the entropy arise from corrections to the saddle point approximation to the integral \eqref{eq:invlap}. 

What's more, the integrals in \eqref{eq:invlap} can be computed in terms of Bessel functions, and one finds that for the class of CFTs considered here, the all order (in the $\frac{1}{\mathcal{E}}$ expansion) re-summed result reads \cite{Loran_2011}: 
\begin{equation}
\rho_{CFT}(\mathfrak{E},\mathfrak{E}) \simeq \frac{\pi^2}{3} c \frac{I_1(S_o(\mathfrak{E}))}{S_o(\mathfrak{E})}+O\left(\exp(-\alpha \mathfrak{E})\right)+O\left(\exp(-\frac{\beta}{\mathfrak{E}})\right).\label{eq:allord}
\end{equation}
Here, $I_1$ is the modified Bessel function of the first kind. 

Therefore, the re-summation of the perturbative corrections to the Cardy formula all come from just the integral \eqref{eq:invlap}. The information of other states in the theory are contained in corrections of the form $\exp(-\alpha \mathfrak{E})$ and $O\left(\exp(-\frac{\beta}{\mathfrak{E}})\right)$. 
\paragraph{The $T\bar{T}+\Lambda_2$ case with $\tau_1=0$:}
A very similar analysis can be carried out to obtain the high energy density of states in the $T\bar{T}$ and $T\bar{T}+\Lambda_2$ deformed CFT, with modular invariance again being the linchpin. The saddle point evaluation of the integral:
\begin{equation}
\rho(\mathfrak{E},\lambda)= \oint d\tau e^{\tau_2 \mathfrak{E}} Z\left(\frac{\lambda}{\tau^2_2},\frac{1}{\tau_2}\right)= \oint d\tau e^{\tau \mathfrak{E}}e^{-\frac{\tau_2}{\lambda}\left(1+\sqrt{\eta+ \frac{c \pi \lambda}{3\tau_2^2}}\right)} \label{eq:ttinvlap}
\end{equation}
gives us, in the $\eta=1$ case: 
\begin{equation}
\rho(\mathfrak{E},\lambda) \simeq \frac{\pi^{3/4}c^{1/4}}{\sqrt{\mathfrak{E}^{3/2}\left(2-\mathfrak{E}\lambda \right)^{3/2}}}e^{\sqrt{\frac{c \pi }{3}\mathfrak{E}\left(2-\mathfrak{E}\lambda\right)}}.
\end{equation}
This result matches that of \cite{Datta:2018thy}, except the sign of $\lambda$ is flipped in comparison with their result. Also, they express their result in terms of dimensionful quantities, whereas the results here are expressed in terms of dimensionless quantities. 

It is interesting to observe that the above equation is simply the rewriting of the Cardy formula for the undeformed CFT density of states:
\begin{equation}
\rho_o(\mathfrak{E})\simeq \frac{c^\frac{1}{4}}{\sqrt{\mathfrak{E}_o^{\frac{3}{2}}}}e^{\sqrt{\frac{2c\pi\mathfrak{E}_o}{3}}},
\end{equation}
but with $\mathcal{E}$ defined through inverting the deformed energy formula:
\begin{equation}
\mathfrak{E} = \left(\frac{1-\sqrt{1-2\mathfrak{E}_o\lambda}}{\lambda}\right),
\end{equation}
i.e.
\begin{equation}
\mathfrak{E}_o = \frac{\mathfrak{E}\left(2-\mathfrak{E}\lambda\right)}{2}. 
\end{equation}
This fact was first noted in \cite{MMV}.

In the $\eta=-1$ case we get:
\begin{equation}
\rho(\mathfrak{E},\lambda) \simeq  \frac{\pi^{3/4}c^{1/4}}{\sqrt{\left(\frac{2-2\mathfrak{E} \lambda +\mathfrak{E}^2\lambda^2}{\lambda}\right)^{\frac{3}{2}}} }e^{\sqrt{\frac{c \pi}{3}\left(\frac{2-2\mathfrak{E}\lambda+\mathfrak{E}^2\lambda^2}{\lambda}\right)}}.
\end{equation}

Here too the sub-leading, perturbative corrections come from the corrections to the saddle point approximation of the integral \eqref{eq:ttinvlap}.

\paragraph{Near Horizon Limit:}
Following the discussion in the previous section, the near horizon limit in the bulk de Siter space corresponds to the value of $\lambda$ where we reach the square root singularity. At the cutoff value of the energy, the following relations hold $\mathfrak{E} =\frac{1}{\lambda_*}=  \frac{c \pi}{3}$. I then evaluate the density of states at this value of $\mathfrak{E}$:
\begin{equation}
\rho(\mathfrak{E},\lambda) \simeq \frac{c}{\left(\frac{c \pi}{3}\right)^{3/2}}e^{\frac{c\pi}{3}}.
\end{equation}
Written in terms of bulk quantities, the Entropy of the cosmological horizon in this ensemble is given by:
\begin{equation}
S_{hor,\tau_1=0} = \frac{\pi \ell}{2G} -\frac{3}{2} \log \left(\frac{\pi \ell}{2G} \right)+\log(\ell/G)+\cdots 
\end{equation}
This result disagrees with \cite{anninos2020quantum} by a factor of two in the logarithmic correction. This is because the result in \cite{anninos2020quantum} is computed in a different ensemble. In particular, to obtain the full Logarithmic contribution, we need to do this calculation in the $\tau_1\neq 0$ ensemble but restrict our attention to the S transformation of the ground state contribution, just like we did in the CFT case. This observation was made in the context of the BTZ black hole entropy in \cite{Sen_2013}.
The analogous saddle point analysis in the case of the $T\bar{T}+\Lambda_2$ deformed theory is very complicated. Instead, I will present an alternative method to obtain the density of states in the deformed theory from a flow equation. 
\subsection{The flow equation for the entropy}
The high temperature limit of the partition function $Z\left(\frac{\lambda}{\tau^2_2},\frac{1}{\tau_2}\right)\bigg\vert_{\tau_2\ll 1}$ involved in the inverse Laplace transform \eqref{eq:ttinvlap} and its generalization to the $\tau,\bar{\tau}\neq 0$ ensemble solves the semiclassical limit of the flow equation. In particular this partition function satisfies the following equation in terms of the $(V,\tau_1,\tau_2)$ variables: 
\begin{equation}
\frac{V(\eta-1)}{4\pi^2 \mu}+\frac{8\pi^2 \mu}{V}\left( \frac{\tau^2_2}{2}\left((\partial_{\tau_1}\ln Z)^2+(\partial_{\tau_2}\ln Z)^2\right)-\frac{V}{2}(\partial_{V}\ln Z)^2\right)-2V \partial_{V}\ln Z=0.\label{eq:hj1}
\end{equation}
This is a rewriting of the Hamilton Jacobi equation that comes from writing \eqref{eq:hc} in terms of $\ln Z = \mathcal{W}+\frac{V}{4\pi^2 \mu}.$ Then, I can rewrite this equation in terms of the $\left(\lambda= \frac{4\pi^2 \mu\tau_2}{V} , \tau_1,\tau_2\right)$ variables. 
Denoting $\ln Z\left(\frac{\lambda}{\tau^2_2},\frac{1}{\tau_2}\right)\bigg\vert_{\tau_2\ll 1} \equiv W$, the equation \eqref{eq:hj1} becomes: 
\begin{equation}
\frac{\tau_2(\eta -1)}{2\lambda}+\frac{\lambda \tau_2}{2}\left((\partial_{\tau_1}W)^2+(\partial_{\tau_2}W)^2\right) +2\lambda \left(1+\lambda \partial_{\tau_2}W\right)\partial_{\lambda}W=0,\label{eq:Weqnt}
\end{equation}
or in terms of the $\tau,\bar{\tau}$ variables: 
\begin{equation}
 -i \frac{(\tau-\bar{\tau})(\eta-1)}{2\lambda}-i\lambda (\tau-\bar{\tau})\left(\partial_{\tau}W\partial_{\bar{\tau}}W\right)+2\lambda\left(1+i\lambda\left(\partial_{\tau}W-\partial_{\bar{\tau}}W\right)\right)\partial_{\lambda}W=0. \label{eq:Weqnttb}
\end{equation}

The inverse Laplace transformation \eqref{eq:ttinvlap} implies that the logarithm of the density of states, the entropy, is related to $W$ by: 
\begin{equation}
S(\mathfrak{E},\bar{\mathfrak{E}},\lambda) = \min_{\tau,\bar{\tau}}\left(i\frac{\tau \mathfrak{E}}{2}-i\frac{\bar{\tau} \bar{\mathfrak{E}}}{2}-W(\tau,\bar{\tau},\lambda)\right).\label{eq:legtrn}
\end{equation}

Then, the equation \eqref{eq:Weqnttb} translates into the following equation for the entropy:
\begin{equation}
\frac{\eta-1}{\lambda^2}(\partial_{\mathfrak{E}}S+\partial_{\bar{\mathfrak{E}}}S)- \mathcal{E}\bar{\mathfrak{E}}(\partial_{\mathfrak{E}}S+\partial_{\bar{\mathfrak{E}}}S)+2\left(-1+\frac{\lambda(\mathfrak{E}+\bar{\mathfrak{E}})}{2}\right)\partial_{\lambda}S=0.
\end{equation}
Here I have made the following replacements that follow from the Legendre transform \eqref{eq:legtrn}:
\begin{equation}
\tau W\rightarrow \partial_{\mathfrak{E}}S,\,\, \bar{\tau}W\rightarrow \partial_{\bar{\mathfrak{E}}}S,
\end{equation}
and 
\begin{equation}
\partial_{\tau}W\rightarrow \mathfrak{E}S,\,\,\partial_{\bar{\tau}}W \rightarrow \bar{\mathfrak{E}}S.
\end{equation}

Recalling that the state of interest has $\mathfrak{E}= \bar{\mathfrak{E}}$, we obtain the equation:
\begin{equation}
\left(-\partial_{\lambda}-\frac{\mathfrak{E}^2}{2}\partial_{\mathfrak{E}}+\lambda \mathfrak{E}\partial_{\lambda}+\frac{(1-\eta)}{2\lambda^2}\partial_{\mathfrak{E}}\right)S(\mathfrak{E},\mathfrak{E},\lambda)=0.\label{eq:entfle}
\end{equation}

Note that this equation can be solved through the method of characteristics. The characteristic equations are:
\begin{equation}
\frac{d\lambda}{dt} = -1+\lambda \mathfrak{E},
\end{equation}
\begin{equation}
\frac{d\mathfrak{E}}{dt} = -\frac{\mathfrak{E}^2}{2}+\frac{1-\eta}{2\lambda^2}.
\end{equation}
which, when combined tell us that:
\begin{equation}
(-1+\lambda \mathfrak{E})\partial_{\lambda}\mathfrak{E}+\frac{\mathfrak{E}^2}{2}+\frac{1-\eta}{2\lambda^2}=0.
\end{equation}
This is the Burgers' equation for $\mathcal{E}$.

The above equation implies that 
\begin{equation}
S(\mathfrak{E},\lambda) = S_{o}\left(\frac{1-\eta-2\mathfrak{E}\lambda+\mathfrak{E}^2 \lambda^2}{\lambda}\right),
\end{equation}

 In the $\eta =1$ case, $S_{o}$ is $S(\mathfrak{E},\mathfrak{E},\lambda=0)$, i.e. the Cardy entropy of the undeformed CFT. This fixes
\begin{equation}
S_{\eta=1}(\mathfrak{E},\mathfrak{E},\lambda) =\sqrt{\frac{c\pi}{3}\mathcal{E}(2-\mathfrak{E}\lambda)}-3\log \left(\sqrt{\frac{c\pi}{3}\mathfrak{E}(2-\mathfrak{E}\lambda)}\right)+\log c+k,
\end{equation}
where $k$ is a constant. The function for the $\eta=-1$ case is fixed by demanding that $S(\mathfrak{E},\mathfrak{E},\lambda\gg 1)\vert_{\eta=1} = S(\mathfrak{E},\mathfrak{E},\lambda\gg 1)\vert_{\eta=-1}$. This gives us the entropy formula 
\begin{equation}
S_{\eta=-1}(\mathfrak{E},\mathfrak{E},\lambda) =\sqrt{\frac{c\pi}{3}\left(\frac{2-2\mathfrak{E}\lambda+\mathfrak{E}^2\lambda^2}{\lambda}\right)}-3\log \left(\sqrt{\frac{c\pi}{3}\left(\frac{2-2\mathfrak{E}\lambda+\mathfrak{E}^2\lambda^2}{\lambda}\right)}\right)+\log c+k',
\end{equation}
where $k'$ is another constant. I can then take the near horizon limit $\mathfrak{E}_* = \frac{1}{\lambda} = \frac{c\pi}{3}$ to find 
\begin{equation}
S_{\eta=-1}\left(\mathfrak{E}_*,\mathfrak{E}_*, \frac{1}{\mathfrak{E}_*}\right) = \frac{c\pi}{3}-3\log \left(\frac{c\pi}{3}\right)+\cdots.
\end{equation}
This is the formula for the entropy of the cosmic horizon including the leading logarithmic correction: 
\begin{equation}
S_{hor} = \frac{\pi \ell}{2G} - 3\log\left(\frac{\pi \ell}{2G}\right)+\cdots.
\end{equation}
The coefficient of the logarithmic term now agrees with the calculation in \cite{anninos2020quantum}, however, the constant term they compute in their one loop calculation is different from the constant that results from the calculation here. It isn't understood why this is the case, and I will leave the investigation of this discrepancy for future work.

\section{Discussion}
The calculation described in the previous section computes the leading high energy density of states along with the sub-leading logarithmic correction by solving the semiclassical flow equation for the entropy. Here, by semiclassical 
I mean a non linear first order equation akin to the Hamilton--Jacobi equation, as opposed to the exact flow equation which is a linear second order equation. The semiclassical flow equation is satisfied only when the contribution of a single energy level is considered. Superpositions of energy levels do not solve this flow equation. However, the sum over all high energy states is related to the $S$ transformation of the ground state- which too satisfies the semiclassical flow equation in the $(\lambda,\tau_a)$ variables. The entropy results from taking the logarithm of the inverse Laplace transformation of the $S$ transformed high temperature partition function, and therefore it too should satisfy the semiclassical flow equation now written in terms of $\mathcal{E},\bar{\mathcal{E}}$ and $\lambda$. The logarithmic correction comes from incorporating quadratic fluctuations around the saddle point evaluation of the inverse Laplace transform, however, it is still just the S transformed ground state contributing to the inverse Laplace transform. This is why the logarithmically corrected entropy still satisfies equation \eqref{eq:entfle}. In fact, all perturbative corrections to the high energy density of states, i.e. corrections involving negative powers of $\mathfrak{E}$ and $\bar{\mathfrak{E}}$ come from the very same inverse Laplace transform.

The contributions from other states in the spectrum, in the CFT calculation appear as corrections of the form $e^{-\frac{\alpha}{\mathfrak{E}}},e^{-\beta\mathfrak{E}}$. 
Therefore, the all order resumed corrections to the Cardy formula in the CFT, \eqref{eq:allord} can be used as initial conditions for equation \eqref{eq:entfle} with $\eta=1$, and then the related solution for $\eta=-1$ can be obtained by demanding matching at large $\lambda$. To support this argument, it would be good to explicitly compute higher order corrections to the saddle point evaluation of the inverse Laplace transform \eqref{eq:ttinvlap}, even in the $\tau_1=0$ ensemble. This exercise will be left to future work. 

Note that the calculation of the density of states presented in this article is strictly in the saddle point approximation, and therefore the methods involved are different from those used in the one dimensional case as presented in \cite{Gross_2020} and \cite{Gross2_2020} where the exact deformed density of states was extracted. It would be interesting to derive the exact deformed density of states by generalising their method to the two dimensional case. 

The linchpin in the above calculation is modular invariance, viewed from the bulk as the large gauge invariance of the theory and from the boundary as being inherited from the seed CFT's modular invariance. However, this means that the spectrum is not truncated to just its real part. The complex energy states necessarily contribute to the sum over high energy states that modular invariance relates to the S transformed ground state. It would be very interesting to better understand the physics of these states and the consequent violation of unitarity. 

An interesting cross check of the results in this article would be to compute the de Sitter entropy along with the subleading corrections from the dS/dS slicing. The de Sitter entropy there has the interpretation of an entanglement entropy between the two matter sectors stitched together at the horizon. Perhaps the functional integral kernel representation of the $T\bar{T}+\Lambda_2$ deformed CFT could be used to this end \cite{Silverstein_Torroba_future}. In fact, the static patch entropy has been computed from a bulk analysis in the context of the dS/dS correspondence in \cite{Geng_2019}. In \cite{Geng_2020}, the holographic entanglement of purification and entanglement wedge cross section have been computed in the dS/dS slicing, and it would be interesting to see whether the cylinder slicing can capture these quantities too. 

The dS/dS correspondence has also been used to argue that de Sitter space is a fast scrambler \cite{Geng_2021} and more recently that this correspondence might provide a resolution of a particular information paradox in de Sitter space \cite{Geng2_2021}. 

In the large $c$/ semiclassical limit, the partition sum representation of the $S^2$ partition function has an entirely real spectrum as discussed in \cite{dSdSTTbar}, \cite{Lewkowycz:2019xse}. This begs the question of whether the unitarity violation witnessed in the static patch slicing is somehow illusory. I will leave investigating these matters to future work. 

The relevance of the cosmic horizon as the locus on which the holographic dual to de Sitter quantum gravity has also been explored in \cite{susskind2021sitter}. It was also argued there that the holographic dual theory to eternal de Sitter is an ensemble average. It indeed would be interesting to connect the $T\bar{T}+\Lambda_2$ deformed $CFT$ to some ensemble averaged theory.

Another approach to study the holographic description of de Sitter space is that of \cite{Banks_2006} (and references therein). This approach identifies matrix like variables for the quantum gravity theory. It would be very interesting to bridge their approach to the deformations of holographic CFTs studied in this work.

\section{Acknowledgements}
I would like to thank Eva Silverstein, Gonzalo Torroba and Victor Gorbenko for their comments on this work. I would also like to thank Leonard Susskind, Pawel Caputa and Zachary Fisher for interesting related discussions. 

\bibliographystyle{utphys}
\bibliography{refs}
\appendix
\section{The ADM Hamiltonian for General Relativity}\label{app:ADM}
The Arnowitt--Deser--Misner (ADM) Hamiltonian and momentum constraints are \cite{Arnowitt:1962hi}
 \begin{equation}
 \mathcal{H}=\frac{1}{\sqrt{g}} g_{i j} g_{k l}\left(\pi^{i k} \pi^{j l}-\pi^{i j} \pi^{k l}\right) - \sqrt{g}(R-2 \Lambda)
\end{equation}
\begin{equation}
\mathcal{H}_{i}=-2\nabla_{j}\pi^{j}_{i}=0.
\end{equation}

The total Hamiltonian can be written as follows:
\begin{equation}
H_{Tot} = \int\textrm{d}^{D}x\left( N(x) \mathcal{H}(x) + \xi^{i}(x) \mathcal{H}_{i}(x)\right)=H(N)+H_{i}(\xi^{i}).
\end{equation}
where the spacetime dimension of the bulk is $D+1$.

We would like to fix the mean curvature of the hypersurface to be constant. To do that we start by
splitting the conjugate momentum $\pi^{ij}$ into traceless and trace components, and define the metric $g_{ij}$ as a conformal rescaling of a constant-curvature counterpart $\bar{g}_{ij}$ via dilaton $\phi(x)$:
\begin{equation} \label{eqn:FieldDecomp}
\pi^{ij}= \sigma^{ij}+\frac{1}{D} \textrm{tr}\pi g^{ij},\,\,\qquad g_{ij}=e^{2\phi(x)}\bar{g}_{ij}.
\end{equation}
The gauge fixing condition imposes the constancy of $T$ defined as:

\begin{equation}
T = \frac{2}{D}\frac{\textrm{tr}\pi}{\sqrt g}, \,\, \nabla_{i}T=0
\end{equation}
Specialising the the case of three dimensional gravity, in this gauge, the Hamiltonian constraint becomes:
\begin{equation}
\mathcal{H}_{CMC}=-\frac{1}{2} \sqrt{\bar{g}} e^{2 \phi}\left(T^{2}-4 \Lambda\right)+\sqrt{\bar{g}} e^{-2 \phi} \sigma^{i j} \sigma_{i j}+2 \sqrt{\bar{g}}\left[\bar{\Delta} \phi-\frac{1}{2} \bar{R}\right]=0,
\end{equation}
where ``barred'' quantities are defined in terms of $\bar{g}_{ij}$. Specialising to $\mathbb{T}^2$, $\bar{g}_{ij}$ depends only on the moduli $\tau_a$ and the conjugate $\sigma^{ij}$ depends only on their conjugates $p^a_{\tau}$. More explicitly: 
\begin{equation}
\bar{g}_{ij} dx^i dx^j = \vert dz +\tau d\bar{z}\vert^2,\,\,z = x^0+ix^1, \bar{z} = x^0-i x^1. \end{equation}
And then, to isolate the conjugate variables to $\tau_a$, we use the following expression \footnote{Given that we identify $p_{\tau_a}=V\frac{\partial_{r}\tau_a}{\tau^2_2}$}: 
\begin{equation}
p^a_\tau= \int d^2x \sigma^{ij} \frac{\partial \bar{g}_{ij}}{\partial \tau^a}
.\end{equation}

Integrating, and changing variables, we find:
\begin{equation}
h_{CMC} = \int \textrm{d}^{D}x\,\,\mathcal{H}_{CMC}=-V^2(T^{2}-4\Lambda) + \tau^{2}_{2}\delta_{ab}p^{a}_{\tau}p^{b}_{\tau}=0.
\end{equation}
Note that classically, the condition for the existence of solutions is $T^{2}\geq 4\Lambda$.  

This Hamiltonian appeared first in \cite{doi:10.1063/1.528475}. For a review of classical and quantum gravity in 2+1 dimensions, see \cite{CarlipNotes}. 

On the reduced phase space, the dynamics of the theory is finite dimensional. It can be quantized as such, and a quantum mechanical theory is obtained. The wavefunctions of interest depend on the modular parameters of the torus, as well as the volume. The corresponding momenta act as derivatives with respect to the conjugate variables: 
\begin{equation}
\hat{T}\psi(V,\tau_{1},\tau_{2})= -\partial_{V}\psi(V,\tau_{1},\tau_{2}),
\end{equation}
\begin{equation}
\hat{p}_{\tau_{a}}\psi(V,\tau_{1},\tau_{2}) = -\partial_{\tau_{a}}\psi(V,\tau_{1},\tau_{2}). 
\end{equation}

With these conventions, the global Hamiltonian constraint equation reads:
\begin{equation}
\hat{h}_{CMC}\,\psi(V,\tau_{1},\tau_{2})= \left[\tau^{2}_{2}(\partial^{2}_{\tau_{1}}+\partial^{2}_{\tau_{2}})-V^{2}\left(\partial^{2}_{V}-4\Lambda\right)\right]\psi(V,\tau_{1},\tau_{2})=0.
\end{equation}

\end{document}